# Electronic Communication Data Link Encryption Simulation Based on Wireless Communication

Rulin Bai

Shijiazhuang Institute of Railway Technology,Shijiazhuang, Hebei ,050041, China

*Corresponding author: Rulin Bai

**Abstract:** In order to improve the simulation effect of electronic communication data link encryption, the author proposes a solution based on wireless communication. The main content of this technology is based on the research of wireless communication, improve the elliptic curve cryptographic algorithm to build a system encryption model, obtain legal and valid node private keys, evaluate and analyze the relevant security attributes of the system, verify the security of the keys, and realize the encryption optimization of wireless network communication. Experimental results show that: Using the improved elliptic curve to simulate the system data chain encryption under the certificateless public key cryptosystem in network communication, the time is only 2.31 milliseconds, which is lower than other algorithms. Conclusion: It is proved that the technology research based on wireless communication can effectively improve the encryption simulation effect of electronic communication data link.

## 1 Introduction

With the continuous development of Internet technology, electronic communication data link technology makes the transmission of information simpler and faster, and can obtain information in real time, transmit information in two directions or multiple directions, analyze and process information, it can also connect multiple platforms together to form a three-dimensional distributed information network, which greatly improves the data transmission speed. However, in this way of information transmission, the security of information transmission has become a problem that people are extremely concerned about at present, especially in the fields of enterprises, troops, hospitals, political and legal institutions, banks, etc., the requirements for information communication security are even more very high [1].

To this end, the world is actively studying the encryption technology of data links, the research found that there are widespread data interception, interruption, eavesdropping, link interference, data tampering and other electronic communication data link security risks. Many scholars around the world have developed a layered security mechanism, the data link is divided into two layers, the access layer and the non-access layer, in order to control the occurrence of the risk of information transmission in the electronic communication data link, but the control is not comprehensive, and the communication data still has the risk of being attacked and eavesdropped. Therefore, some scholars propose that in the data link, increase the interference code, improve the complexity of the electronic communication signal, and avoid the eavesdropper from eavesdropping on the communication. However, this method also reduces the utilization rate of the communication line [2].

The data link has the characteristics of high communication transmission efficiency, strong anti-interference ability and good confidentiality. The increasing complexity of communication systems increases the cost of system research and design, in order to better study communication systems, it is necessary to establish an effective system evaluation model and design a reasonable evaluation method. Through the research on system modeling and simulation technology, the system evaluation work will be further carried out. Whether it is to improve the existing system or design a new system, or to evaluate and analyze the relevant security attributes of the system, verify the security of the key, improve the encryption technology of electronic communication data chain, and resist the risk of information transmission, it has important reference value and guiding significance to realize the encryption optimization of large-scale network communication. The data link simulation model is shown in Figure 1.

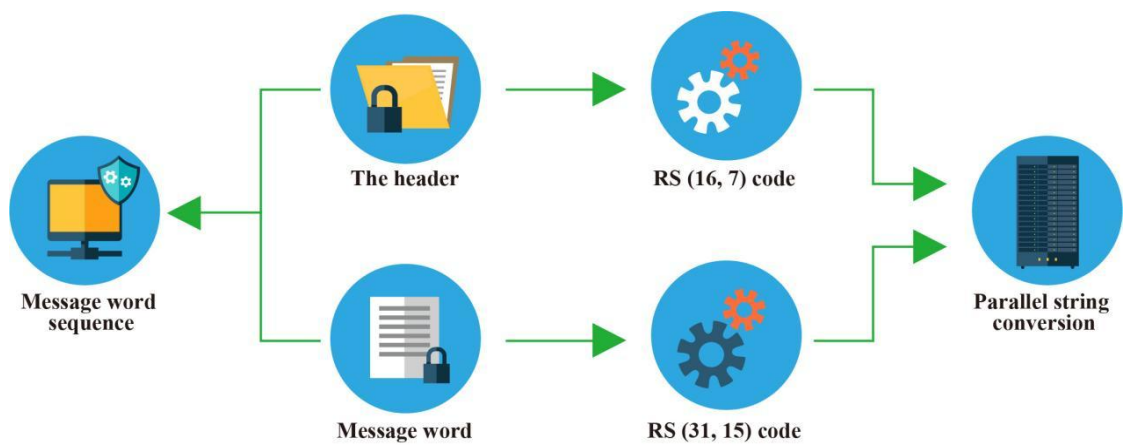

**Figure 1 Data link simulation model**

## 2 Literature review

The emergence of wireless communication technology has made progress in communication technology, freed human communication from the constraints of time, space and products, improved people's quality of life, and accelerated the process of social development.Lee, S. et al. It means that wireless communication technology is developing in the direction of broadband and multimedia data services, and finally realizes all kinds of communication with all products anytime, anywhere [3]. The future wireless communication will be integrated with the Internet, but from fixed access to the Internet to wireless mobile access to the Internet, while wireless network technology brings great convenience to people, security issues have become a major obstacle to the popularization of wireless network technology applications. The security mechanism of the Internet itself is relatively fragile. The inherent openness of Internet wireless network transmission media and the limitations of mobile device storage resources and computing resources, therefore, in the wireless network environment, not only all the security threats in the wired network environment, but also the emerging security threats specifically for the wireless environment must be faced. Therefore, when building a wireless network system, in addition to providing a complete multi-service platform on the wireless transmission channel, the encryption security of the wireless network system must also be considered. When encrypting network connections, all data except valid and invalid nodes can be encrypted to provide cryptographic complexity. Wang, J. et al. When the network is large and there are a large number of invalid nodes, traditional symmetric cryptography says that if there are too many nodes to communicate securely, all nodes need to exchange keys securely [4]. Large networks have the disadvantages of slow decryption and high value. At present, the encryption methods in the form of non-certified public keys in the public communication network include encryption methods based on DES algorithm, encryption methods based on digital signature algorithm and encryption process based on IDEA algorithm. Among them, the most used is the encryption party in the public key cryptosystem based on the digital signature algorithm. Since the effective encryption of large-scale network communication can ensure communication security, it plays an irreplaceable role in communication security. Therefore, it has become one of the key research topics of relevant scholars and has a very broad development prospect. In large network communications. There are major defects in the encryption process under the certificateless public key cryptosystem, when the traditional certificate encryption communication algorithm encrypts network communication, in order to ensure the complexity of encryption, only all data can be encrypted, without considering the validity and Uselessness. When the network scale is large and the number of nodes is large, before traditional symmetric cryptography can communicate securely, all nodes must exchange keys to ensure security. Yang, C. Y. believes that it will cause the disadvantages of slow encryption and decryption and large key size [5].

The purpose of the above problems, the encryption optimization algorithm in large-scale communication network based on wireless network and proofless public key cryptosystem is proposed, and fixed elliptic curve cryptography is developed. Yin, S. Carry out system initialization according to relevant theories, establish a system encryption model according to the improved elliptic curve cryptographic algorithm, obtain legal and valid node private keys, and evaluate and analyze relevant security features. Verify the security of systems and keys and implement electronic encryption optimization of telecommunications data networks [6].

## 3 Methods

### 3.1 Wireless Communication Technology

(1) UWB ultra-wideband wireless access technology

Wireless communication uses electromagnetic waves to transmit signals to achieve information exchange, and it also has many characteristics in the form of communication. UWB ultra-wideband wireless access technology is built on the basis of pulse radio technology, this technology not only eliminates the carrier form that cannot be solved by traditional radio technology, but also consumes relatively low energy and cost, and this technology also has many advantages, high transmission efficiency, strong anti-interference and so on.

(2) RFID automatic identification technology

Compared with other communication technologies, this technology can avoid more manual intervention, and it can also overcome the impact of harsh environments, and the operation of communication technology will be more convenient [7]. Today, RFID automatic identification technology has been widely used in China's logistics and communications industries [8].

(3) WIFI wireless network technology

**Table 1 Advantages of wireless communication technology**

| type of technology | Application advantages |
|---|---|
| UWB ultra-wideband wireless access technology | Good confidentiality performance, strong anti-interference ability, high transmission efficiency |
| RFID automatic identification technology | It has strong adaptability, simple and convenient technical operation, and high efficiency |
| WIFI wireless network technology | Wireless communication has high processing efficiency, can save a lot of operating costs, and is convenient and fast for people to use |

With the development of social science and technology, the wireless network communication technology has been continuously improved, especially the WIFI wireless network communication technology, which has covered a full range of people's life and work, this not only improves people's lives, but also increases their productivity. Moreover, the technology can also be connected to smart devices, which not only facilitates people, but also greatly improves the efficiency of wireless communication processing. Table 1 shows the relative advantages of the three wireless communication technologies [9].

**3.2 Development Status of Wireless Communication Technology**

(1) 5G technology

With the development and application of 3G and 4G technologies, the communication technology in the post-4G era is called 5G communication technology. Initially, Japan's NTT company proposed 5G communication, and China proposed the concept of 5G at the International Communication Conference. Although the current 5G communication technology standards have not been unified, it has brought great convenience to people's production and life [10].

(2) Satellite communication technology

Satellite communication technology can realize wireless communication, especially in areas where users are very sparsely distributed, the use of satellite communication technology can enable user equipment to connect to wireless communication, if connected to a wired network, the safety and efficiency of communication will be greatly improved, but this technology will incur a lot of operating costs during application. Furthermore, there are few satellite platforms on the ground, which limits the application of satellite communication technology to a large extent [11].

(3) Bluetooth technology

For public wireless networks with a short distance, access points can be used to transmit information and data, which can largely prevent the complexity of cable laying, especially for teleconferencing and other work, which can be achieved with the help of bluetooth technology, and with modern technology with the rapid development of the technology, the application scope of this technology is also expanding [12].

(4) Wireless broadband technology

With the advancement of science and technology, the access methods of wireless broadband have also increased. The first is microwave broadband access technology, when applying this technology, the frequency band needs to be controlled at around 28GHz, the "cellular" network layout can also effectively reduce the loss caused by the transmission distance, and the "cellular" network layout can also reduce the transmission power of wireless communication, so as to achieve short-range bidirectional image and data transmission [13].

The second is satellite access technology, which is widely used in real estate, finance and education industries, under the function of this technology, it can not only quickly access the Internet, but also realize the effective distribution of data packets, because of the strong stability of this technology, it is favored by various industries. Furthermore, the infrared optical communication access technology, the transmission speed of this technology during the application period is extremely fast, usually 4-619MB/s, which can quickly transmit data information, and is affected by the infrared working band, the distance of information transmission is as high as hundreds of meters, and it will not interfere with other communication systems during the transmission process, when transmitting and receiving wireless signals, optical instruments are used. Finally, there is the multi-point microwave access technology, which is suitable for many low-frequency bands, and only applies to the three bands of 5.8, 2.5, and 3.5 GHz. Due to the limitations of this technology, its application range is small [14].

(5) Ultra-broadband technology

Ultra-wideband technology is based on the wireless carrier with the help of nanosecond non-sinusoidal narrow wave, data transmission in the form of pulse, coupled with the wide coverage spectrum of ultra-wideband technology, so in the case of low power consumption and low program, data transfer is also possible [15].

(6) WIFI technology

The technology is suitable for office and family activities, the coverage of WIFI can reach about 90 meters, and the transmission speed of the signal is extremely fast, moreover, the technology can be formed into a wireless local area network, so that offices on the same floor can use this network for office work. WIFI technology can also help users access Web and email, and provide users with faster and more convenient network services at home and on the go, as an extension of high-speed wired access technology, WIFI technology has the characteristics of mobility and low price, it has been widely used in many fields where wired access needs to be wirelessly extended. In addition, due to the great difference in data rate and coverage, in the field of broadband applications, WIFI technology has become a supplement to wired access technology, and in terms of telecom operators, the positioning of this technology will gradually change to cellular mobile communication Supplement [16].

### 3.3 The principle of encryption in network communication

In the case of the communication algorithm encrypted by the communication network encryption certificate, in order to ensure the complexity of encryption, all data must be encrypted, and all nodes must be exchanged securely, the principle of the encryption process. As follows:

In large-scale network communication, the algebraic closed field representing the field K is set to be $\bar{K}$, and $K^* = K/\{0\}$ is a multiplicative group, and its constituent elements are all non-zero elements in K. If the affine plane of the algebraic closed field $\bar{K}$ is $\bar{K} \times \bar{K}$, namely: $A^2(\bar{K})$, then

$$A^2(\bar{K}) = \bar{K} \times \bar{K} = \{(x,y): x,y \in \bar{K}\} \tag{1}$$

It is known that the affine plane curve on the closed algebraic field $\bar{K}$ is a set composed of zeros in the polynomial $C \in \bar{K}[X,Y]$, that is,

$$C = \{(x,y) \in A^2(\bar{K}): C(x,y) = 0\} \tag{2}$$

It is known that C is an affine plane curve, $P = (x,y)$ is a point on this curve, if the conditions are met

$$\frac{\partial C}{\partial X}(x,y) = \frac{\partial C}{\partial Y}(x,y) = 0 \tag{3}$$

Then $P = (x, y)$ is the singular point on C, then the curve C is a singular curve, on the contrary, if the conditions of any point on the curve are not satisfied, the curve is a non-singular curve.

If E satisfies the following equation

$$E: y^2 + a_1 + xy + a_3y = x^3 + a_2x + a_4x + a_6 \tag{4}$$

Then it is an elliptic curve, which can be described by the following formula

$$E = \{(x, y) \in A^2(\bar{K}) : E(x, y) = 0\} \cup \{O\} \tag{5}$$

Assuming that $P = (x, y)$ is a point on E and satisfies $x, y \in K$, then P is a rational point of K. Then the point group $E(K)$ composed of points on the elliptic curve can be described by the following formula:

$$E(K) = \{(x, y): x, y \in K, E(x, y) = 0\} \cup \{O\} \tag{6}$$

Set up

$$b^2 = a_1^2 + 4a_2, b_4 = 2a_4 + a_1a_3, \\ b_6 = a_3^2 + 4a_6 \tag{7}$$

$$b_8 = a_1^2a_6 + 4a_2a_6 - a_1a_2a_4 + a_2a_3^2 - a_4^2, \\ c_4 = b_2^2 - 24b_4 \tag{8}$$

Then the discriminant formula of elliptic curve E in large network communication can be described as follows

$$\delta(E) = -b_2^2b_8 - b_4^3 - 27b_6^2 + 9b_2b_4b_6 \tag{9}$$

In the formula, only when $\delta = 0$, the curve is a singular curve.

When E is isomorphic, its properties remain unchanged, so its simple type can be used for calculation, which is a standard encryption process, which can be described as follows:

1) If

$$\text{Char }(K) \neq 2,3 \text{ then: } y^2 = x^3 + a_4x + a_6 \tag{10}$$

2) If Char $(K) = 2$, and $j \neq 0$ then

$$y^2 + xy = x^3 + a_2x^2 + a_6 \tag{11}$$

3) If Char $(K) = 2$, and $j = 0$ then

$$y^2 + a_3y = x^3 + a_4x + a_6 \tag{12}$$

4) If AA, and BB then

$$y^2 = x^3 + x^2 + a_6 \tag{13}$$

According to the principle described above, the encryption processing under the certificateless public key system in large-scale network communication can be realized, and the security requirements of large-scale network communication can be met [17].

**3.4 Encryption optimization under certificateless public key cryptosystem**

In large-scale network communication, in the process of certificate encryption without public key cryptosystem, due to the huge network scale, the calculation value of using traditional algorithms is too high, the computer efficiency is very low, and the system is not. Therefore, based on the development of elliptic curve cryptography, the author proposes an encryption optimization algorithm in a certificateless public key cryptosystem for large-scale network communication [18].

（1）Initialize the communication system

Initializing the system provides data support for the construction of the encryption model. The initialization method is as follows:

1) Randomly select a prime number p of length $k - bit$ and the elliptic curve parameter group denoted by $\{F_p, E(F_p), G\}$, among them, $F_p$ is a finite field composed of p elements, $E(F_p)$ is the set of all rational points contained on the elliptic curve E on $F_p$, G is one of the base points on E;

2) Randomly select $x \in Z_n^*$, and obtain the value of $P = x \cdot G$ through calculation, among them, x is used to represent the master key of the system, and P is used to represent the public parameters of the system corresponding to x;

3) Select two password hash functions: $H_1:\{0,1\}^* \to Z_n^*$、$H_2:\{0,1\}^* \to Z_n^*$ ;

4) Determine the system parameters disclosed by the system as $params = \{F_p, E(F_P), G, P, H_1, H_2\}$;

Choose any of the functional formulas $f(x) = s + \sum_{i=1}^{t-1} a_i x^i (\mod q)$ and $f(0) = s$ . Randomly select n communication nodes $ID_V$ in large-scale network communication as pre-selected nodes, obtain the master key component $s_V = f(ID_V) \mod q$ through calculation and distribute it, and verify the parameter $W_s^V = s_V P \in G_1$.

In the process of initializing the system, the $KGC$ Trusted Key Generation Center) generates part of the private key $D_V = sH_1(ID_V//phase_0) = sQ_V$ for the node, assuming that the secret value selected by node $ID_V$ is $x_V \in Z_q^*$, the complete private key of $ID_V$ is $SK_V = x_V D_V$, and its public key is $PK_V = x_V P_{pab}$.

(2) Building a system encryption model

The system encryption model is constructed according to the relevant theory of elliptic curve cryptography, the steps are as follows:

Step1: Node $ID_A$ selects any number of $x_A \subset Z_q^*$、 $\subset Z_q^*$, and performs related calculations on $PK = x_A P, R = \tau P$;

Step 2: The node $ID_A$ calculates $Q_A = H_1(ID_A//phase_i)$, and the public key is known to be $PK_A = x_A P_{pub}$, then broadcasts the private key update request related message $R_{update} = \{Q_A, R, PK, PK_A, ID_A\}$ to the node set V.

Step 3: After receiving the above request of the previous option, first determine whether the node is deleted. If it is deleted, the process ends; if not, please check the validity of the node, and the check formula is $\hat{e}(PK_A, P) = \hat{e}(PK, P_{pub})$, if the verification result does not hold, the algorithm ends. Calculate part of the private key $m_A = s_x Q_A$ of node $ID_A$, where $s_x$ represents master key sharing.

Step4: $ID_X$ adds the above request message, randomly selects the random number $\tau_u$, and returns the verification parameter $W_v^x$ and the relevant part of the update response message ciphertext pair:

$$\varepsilon_A = \langle U, V \rangle = \langle m_A + \tau_v R, \tau_v P \rangle \tag{14}$$

Step5: $ID_A$ calculates the private key message issued by $ID_X$ according to the received response message:

$$m_A = U - \tau V = s_x Q_A \tag{15}$$

According to the verification parameter $W_s^x$, $\hat{e}(Q_A, W_s^x) = \hat{e}(s_x Q_A, p)$ is verified, if it is established, part of the private key message is accepted, on the contrary, part of the message is considered illegal.

Step6: Reconstruct part of the private key $D_A = \Sigma \lambda_X(0) s_v Q_A = s Q_A$, and calculate the private key $SK_A = x D_A$ according to the node $ID_A$.

(3) Analysis of Key Security

According to the relevant theory, the relevant security properties of the system after the encryption and optimization of the improved algorithm are evaluated and analyzed, the content of the evaluation is as follows:

1) Security evaluation of known session keys During the execution of the system, a new session key will be generated based on different temporary values for each new session. Therefore, when one of the keys is compromised, there will be no impact on the security of other sessions [19].

2) Analysis of Forward Security

For the first type of attack, $I_2(\hat{e}(SK_A, T_B))$ can be obtained through operations based on the private key of node $ID_A$, but since the private key and b of node $ID_B$ and a are unknown, therefore, $H_2(\hat{e}(aQ_B, PK_B))$ cannot be calculated: For the second type of attack, if the master key of the system is known, part of the private keys $D_A$ and $D_B$ can be obtained by calculation, however, since $x_A, x_B$ is not aware of it, the values of $H_2(\hat{e}(SK_d, T_B))$ and $H_2(\hat{e}(aQ_B, PK_B))$ cannot be obtained; It is assumed that both types of attacks have obtained the long-term private key of the system, but since $a, b$ has no knowledge of it, $H_2(\hat{e}(aQ_B, PK_B))$ cannot be calculated [20].

3) Analysis of key security

If the first type of attack knows the private key $SK_A$ of $ID_A$ and wants to replace $ID_B$, after obtaining $T_B$ according to $SK_A$, $H_2(\hat{e}(SK_A, T_B))$ can be calculated, but $H_2(\hat{e}(aQ_B, PK_B))$ cannot be calculated; In the second type of attack, part of the private key $D_A, D_B$ can be obtained based on the master key, but $H_2(\hat{e}(aQ_B, PK_B))$ cannot be calculated.

4) Unknown key sharing

Therefore, such attacks must know the user's long-term private key, which is difficult to implement [21].

From the above analysis, it can be seen that, using the encryption optimization algorithm under the certificateless public key cryptosystem in large-scale network communication based on elliptic curve cryptography, its security can be guaranteed [22].

**4 Results and Analysis**

Experiments are required to verify the effectiveness of the developed algorithm. During the experiment, the performance of the improved algorithm was compared with the traditional DES algorithm, digital signature algorithm and IDEA algorithm encryption method [23]. The MIRACAL

[SSLM] standard cryptographic tool was successfully used in the calculation, the CPU was CPIV3-GHz, the memory was 1G, and the operating system was the Windows XP hardware platform [24]. In order to achieve the preset security level, an elliptic curve $E(F_p): y^2 = x^3 + x$, with a penetration degree of 2 is used, where p represents a prime number that satisfies the condition $p + 1 = 12qr$ [25].

In Figure 2, P and Q represent the assumed rational points and the two points are not different, 1 is a straight line passing through these two points, R represents the third intersection of 1 and the curve, then point G is G=P+Q , the improved elliptic curve algorithm based on this can be shown in Figure 3. As shown in Figure 2 and Figure 3.

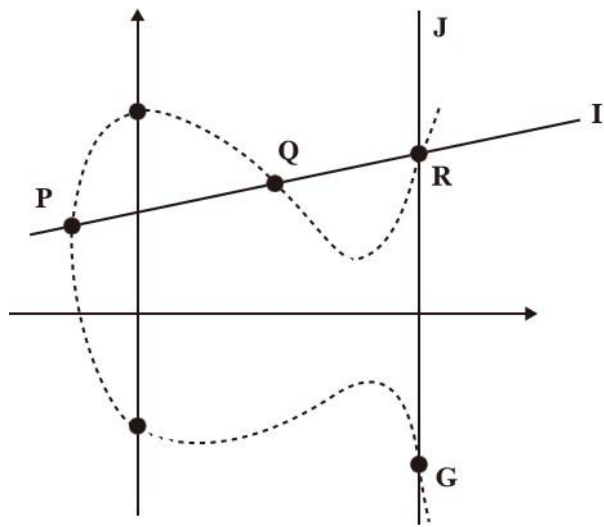

Figure 2 Elliptic curve plus group algorithm

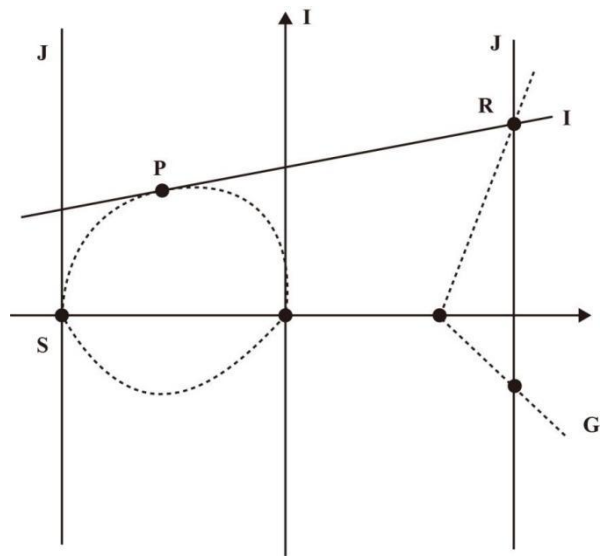

**Figure 3 Improved elliptic curve plus group algorithm**

As shown in Figures 2 and 3 above, the encrypted communication network improved by the elliptic curve algorithm has a higher security factor. During the experiment, the performance of different encryption algorithms was compared, and the results are shown in Figure 4.

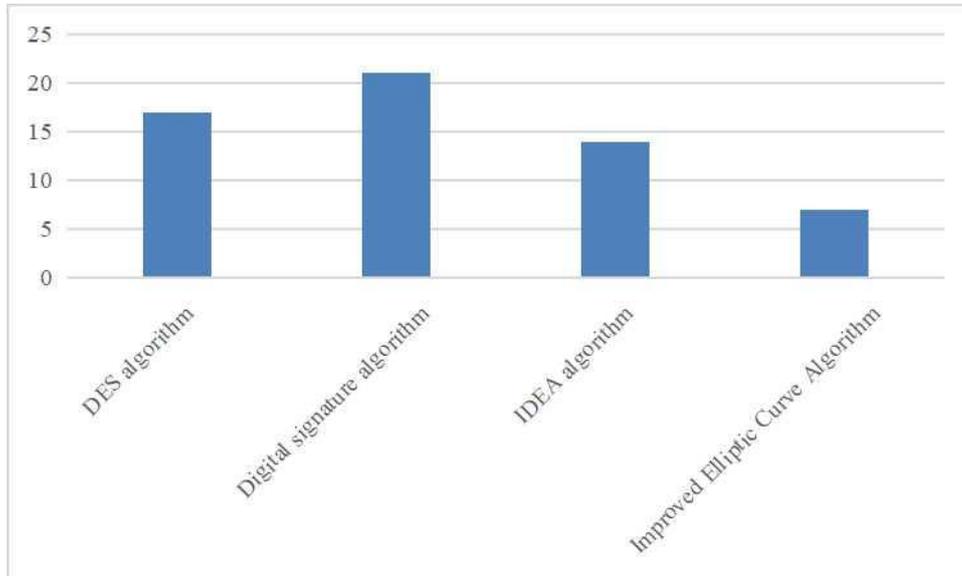

**Figure 4 Operation time of different algorithm units**

Through the analysis of the test results, the test results shown in Table 2 can be obtained.

**Table 2 Operation time of basic unit of different algorithms**

| algorithm | DES algorithm | Signature algorithm | IDEA algorithm | improve algorithm |
| --- | --- | --- | --- | --- |
| Time (milliseconds) | 16.25 | 20.42 | 9.76 | 2.31 |

According to the above test, the elliptic curve encryption certificate public key cryptosystem developed in the communication network only needs 2.31 milliseconds to use the data chain, which is lower than the other three algorithms and has great advantages in encryption efficiency.

**5 Conclusion**

The author introduces the technology based on wireless communication, develops the elliptic curve cryptographic algorithm, establishes the encryption model, obtains legal and effective node privacy, measures and analyzes the security characteristics of the system, and verifies the security of the machine. Importance, knowledge of electronic communication, cryptographic optimization of chains. By developing elliptic curve, the method encrypts data connection in the form of public key in network communication, and compares it with the encryption methods of basic DES algorithm, digital signature algorithm and IDEA algorithm, and draws the conclusion as follows: elliptic curve algorithm has only 2.31 milliseconds, better than the above three less than algorithms.Therefore, it is proved that the technical research based on wireless communication can effectively improve the encryption simulation effect of electronic communication data link.

**References**

[1] Verkhovsky, B. S. . (2011). Information protection based on extraction of square roots of gaussian integers. International Journal of Communications, Network and System Sciences, 4(3), 133-138.

[2] Sridhar, J. , &  Lingeswaran, G. . (2021). A novel approach on plcc based domestic data communication. IOP Conference Series: Materials Science and Engineering, 1084(1), 012092-.

[3] Hidayat, R. ,  Reza, F. , Herawati,  Afiyah, S. ,  Lestari, N. S. , &  Mahardika, A. G. , et al. (2021). Remote monitoring application for automatic power supply system in telecommunication network. Journal of Physics: Conference Series, 1933(1), 012103-.

[4] Wang, J. ,  Liang, Y. , &  Wanga, G. . (2021). The one-dimensional liquid infiltration characteristics of ionized rare earth based on wireless network sensor detection system. IEEE Sensors Journal, PP(99), 1-1.

[5] Lateef, Z. A. ,  Al-Azzwi, F. F. , &  Ibrahim, M. S. . (2021). Design and performance thulium doped fiber amplifier in optical telecommunication networks. Journal of Physics: Conference Series, 1963(1), 012006-.

[6] Yin, S. ,  Liu, J. , &  Teng, L. . (2020). Improved elliptic curve cryptography with homomorphic encryption for medical image encryption. International Journal of Network Security, 22(3), 421-426.

[7] Mahmood, S. N. ,  Ishak, A. J. ,  Ismail, A. ,  Soh, A. C. , &  Alani, S. . (2020). On-off body ultra-wideband (uwb) antenna for wireless body area networks (wban): a review. IEEE Access, PP(99), 1-1.

[8] Fujisaki, K. . (2020). Performance evaluation of table type rfid reader for library automatic book identification. International journal of web information systems, 16(1), 65-78.

[9] Feng, T. , Yue, Y. , Xu, Y. , Lyu, B. , & Guan, G. . (2020). Predicted decoupling for coexistence between wifi and lte in unlicensed band. IEEE Transactions on Vehicular Technology, 69(4), 4130-4141.

[10] Dai, X. , Debregeas, H. , Rold, G. D. , Carrara, D. , & Lelarge, F. . (2021). Versatile externally modulated lasers technology for multiple telecommunication applications. IEEE Journal of Selected Topics in Quantum Electronics, 27(3), 1-12.

[11] Jiang, J. , Zuo, J. , & Feng, H. . (2022). Research on equalization technology of broadband satellite communication channel. Journal of Physics: Conference Series, 2209(1), 012005-.

[12] Huang, Y. C. , Parimi, V. , Chang, W. C. , Syu, H. J. , & Lin, C. F. . (2021). Silicon-based photodetector for infrared telecommunication applications. IEEE Photonics Journal, PP(99), 1-1.

[13] Liu, Y. , Li, H. , & Zhang, M. . (2021). Wireless battery-free broad-band sensor for wearable multiple physiological measurement. ACS Applied Electronic Materials, 3(4), 1681-1690.

[14]ED Santis, A Giuseppi, A Pietrabissa, M Capponi, & FD Priscoli. (2022). Satellite integration into 5g: deep reinforcement learning for network selection. Machine Intelligence Research, 19(2), 127-137.

[15] Khaliev, S. U. , Bolotnev, A. Y. , & Ignatev, I. V. . (2021). Types of modulation in ir ultra-wideband technology. Journal of Physics: Conference Series, 2032(1), 012010-.

[16] Xu, N. , Song, Y. , & Meng, Q. . (2021). Application rfid and wi-fi technology in design of iot sensor terminal. Journal of Physics: Conference Series, 1982(1), 012182 (8pp).

[17] Guo, Y. , Li, J. , Jiang, M. , Yu, L. , & Wei, S. . (2020). Certificate-based encryption resilient to continual leakage in the standard model. Security and Communication Networks, 2020(7), 1-11.

[18] Koirala, K. P. , Garcia, H. , Sandireddy, V. P. , Kalyanaraman, R. , & Duscher, G. . (2021). Bimetallic fe–ag nanopyramid arrays for optical communication applications. ACS Applied Nano Materials, 4(6), 5758-5767.

[19] Wu, L. , Zhou, J. , & Li, Z. . (2020). Applying of ga-bp neural network in the land ecological security evaluation. IAENG Internaitonal journal of computer science, 47(1), 11-18.

[20] Vishnevsky, V. M. , Morozov, V. P. , & Alikin, K. A. . (2021). Challenges in design of an on-board power supply system for a tethered unmanned aerial telecommunication platform. Journal of Physics: Conference Series, 2091(1), 012034-.

[21] Yuvaraj, N. , Srihari, K. , Dhiman, G. , Somasundaram, K. , & Masud, M. . (2021). Nature-inspired-based approach for automated cyberbullying classification on multimedia social networking. Mathematical Problems in Engineering, 2021, 1-12.

[22] Selva, Deepaa & Pelusi, Danil & Rajendran, Arunkumar & Nair, Ajay. (2021). Intelligent Network Intrusion Prevention Feature Collection and Classification Algorithms. Algorithms. 14. 224.

[23] Chen, J. , Liu, J. , X Liu, X Xu, & Zhong, F. . (2020). Decomposition of toluene with a combined plasma photolysis (cpp) reactor: influence of uv irradiation and byproduct analysis. Plasma Chemistry and Plasma Processing.

[24] Huang, R., Zhang, S., Zhang, W., Yang, X. Progress of zinc oxide-based nanocomposites in the textile industry, IET Collaborative Intelligent Manufacturing, 2021, 3(3), pp. 281–289.

[25] Zhang, Q. (2021). Relay vibration protection simulation experimental platform based on signal reconstruction of MATLAB software. Nonlinear Engineering, 10(1), 461-468.